\documentclass{article}

\usepackage[utf8]{inputenc} 
\usepackage[T1]{fontenc}    
\usepackage{hyperref}       
\usepackage{url}            
\usepackage{booktabs}       
\usepackage{amsfonts}       
\usepackage{nicefrac}       
\usepackage{microtype}      
\usepackage{amsmath}
\usepackage{graphicx}
\usepackage{wrapfig}
\usepackage{caption}

\title{Efficient ConvNets for Analog Arrays}

\author{
  Malte J. Rasch \and Tayfun Gokmen \and Mattia Rigotti \and Wilfried Haensch \\[0.5cm]
  IBM Research AI\\
  TJ Watson Research Center\\
  Yorktown Heights, NY 10592
}
\begin{document}

\maketitle

\newcommand{\np}{{n_p}}
\newcommand{\nh}{h}
\newcommand{\nw}{w}
\newcommand{\kw}{k_w}
\newcommand{\kh}{k_h}
\newcommand{\nc}{c_\text{in}}
\newcommand{\ncout}{c_\text{out}}
\renewcommand{\k}{k}
\newcommand{\K}{K}
\newcommand{\I}{I}
\renewcommand{\a}{\mathbf{b}}
\newcommand{\M}{M}
\newcommand{\ntiles}{n_t}

\renewcommand{\eqref}[1]{Eq.~\ref{eq:#1}}
\newcommand{\figref}[1]{Fig.~\ref{fig:#1}}
\newcommand{\tabref}[1]{Tab.~\ref{tab:#1}}
\newcommand{\secref}[1]{Sec.~\ref{sec:#1}}

\newcommand{\lr}{\lambda}
\newcommand{\lrgamma}{\lambda_\gamma}

\newcommand{\remark}[1]{{ \bf [ \footnotesize #1 ]}}
\newcommand{\train}[1]{{\tiny (#1)}}

\begin{abstract}
  Analog arrays are a promising upcoming hardware technology with the
  potential to drastically speed up deep learning.  Their main
  advantage is that they compute matrix-vector products in constant
  time, irrespective of the size of the matrix.  However, early
  convolution layers in ConvNets map very unfavorably onto analog
  arrays, because kernel matrices are typically small and the constant
  time operation needs to be sequentially iterated a large number of
  times, reducing the speed up advantage for ConvNets. Here, we
  propose to replicate the kernel matrix of a convolution layer on
  distinct analog arrays, and randomly divide parts of the compute
  among them, so that multiple kernel matrices are trained in
  parallel. With this modification, analog arrays execute ConvNets
  with an acceleration factor that is proportional to the number of
  kernel matrices used per layer (here tested 16-128).  Despite having
  more free parameters, we show analytically and in numerical
  experiments that this convolution architecture is self-regularizing
  and implicitly learns similar filters across arrays.  We also report
  superior performance on a number of datasets and increased
  robustness to adversarial attacks.  Our investigation suggests to
  revise the notion that mixed analog-digital hardware is not suitable
  for ConvNets.
\end{abstract}

\section{Introduction}

Training deep networks is notoriously computationally intensive.  The
popularity of ConvNets is largely due to the reduced
computational burden they allow thanks to their parsimonious number of
free parameters (as compared to fully connected networks), and their
favorable mapping on existing graphic processing units (GPUs,
\cite{chetlur2014cudnn}).

Recently, speedup strategies of the matrix multiply-and-accumulate
(MAC) operation (the computational workhorse of deep learning) based
on mixed analog-digital approaches has been gaining increasing
attention.  Analog arrays of non-volatile memory provide an in-memory
compute solution for deep learning that keeps the weights
stationary~\cite{yang2013memristive, fumarola2016accelerating}.
As a result, the forward, backward and update steps of
back-propagation algorithms can be performed with significantly reduced data
movement. In general, these analog arrays rely on the idea of
implementing matrix-vector multiplications on an array of analog
devices by exploiting their Ohmic properties, resulting in a one-step
constant time operation, i.e. with execution time {\it independent} of
the matrix size (up to size limitations due to the device
technology)~\cite{gokmen2016acceleration}.

Matrix-matrix multiplications can harness this time advantage from
analog arrays, but since they are implemented as a sequence of
matrix-vector products, their execution time is proportional to the
number of such products.  In other words, the time required to
multiply a matrix on an analog array of size $n_o\times n_s$ with an
input matrix of size $n_s\times n_p$ is not proportional to the
overall amount of compute ($\propto n_on_s\np$, as for conventional
hardware~\cite{he2015convolutional}), but instead only scales linearly
with the number of columns of the input matrix $\np$ and is invariant
with respect to the size of the matrix stored on the analog array
($n_o\times n_s$).

These considerations indicate that ConvNets do not map favorably onto
analog arrays~\cite{gokmen2017training}, as becomes clear when one
formulates the convolution operation in terms of a matrix-matrix
product (see \secref{tile_methods} for a detailed derivation). It
turns out that kernel matrices (obtained by flattening and stacking
convolution filters), are typically small, corresponding to a small
size of the analog $n_o\times n_s$-array. More crucially,
matrix-vector products need to be iterated $\np$ times (the number of
image patches), which is proportional to the total number of pixels in
the input image and can thus be very large, particularly for early
conv layers.

A common strategy to speed up training is to use data parallelism,
where updates over large batches of data are computed in parallel on
independent computing nodes and then averaged
(e.g.\cite{you2017imagenet}).  However, this is not a practical
solution to speed up training on analog arrays, since weight updates
are computed only implicitly on stationary weights in non-volatile
memory and are thus not directly accessible for
averaging~\cite{gokmen2016acceleration}.

\begin{figure}[t!]
  \centering
  \includegraphics[width=1\textwidth,clip, trim=0cm 0cm 0cm 0cm]{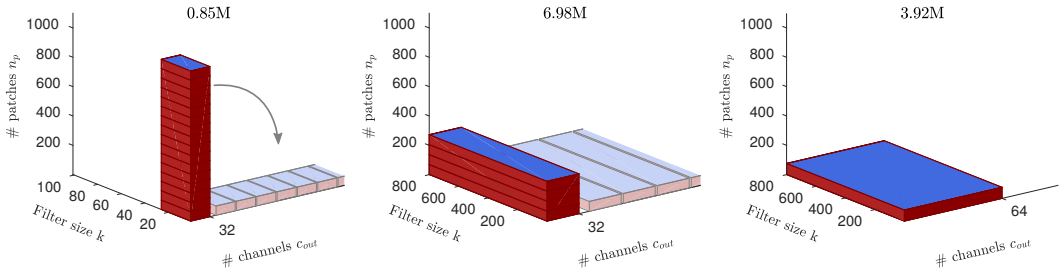}
  \caption{The amount of compute for the example ConvNet (respective
    for the 3 layers). Blue areas ($\k \times \ncout$) indicate the
    size of the kernel matrices.  Computing time for analog arrays is
    proportional only to $\np$ and peaks at the first layer, while the
    amount of compute is $O(\np\k\ncout)$ (the volume of the red
    cuboid; MACs in titles) and peaks at the second layer. For each
    layer, our approach distributes the compute onto multiple replica
    of the kernel matrix residing on distinct arrays (``tiles''),
    indicated as tilings of the red cuboids into $\ntiles=(16,4,1)$
    small boxes, respectively. Since tiles are trained independently
    and in parallel, the compute time on analog arrays effectively
    becomes constant across layers (same height across layers; note,
    however, that the number of output channels of the convolution
    does not change). Our tiling schemes refer to the way individual
    image patches are assigned to the tiles.}
  \label{fig:compute_volume}
\end{figure}

Here, we propose a simple solution to accelerate ConvNets on analog
arrays, which we call RAPA Convolution (for {\it Replicated Arrays
  with Permuted Assignment}).  The main idea is to use model
parallelism to reduce the overall computation {\it time} on analog
arrays (but not the {\it amount} of computation, as done e.g.
in~\cite{figurnov2016perforatedcnns}).  Concretely, we propose to
replicate the kernel matrix onto $\ntiles$ separate analog arrays
(``tiles''), and to distribute the compute equally among the tiles
(see \figref{compute_volume}).  When this architecture proposed for
analog arrays is simulated on conventional hardware (as we do here),
it is equivalent to learning multiple kernel matrices independently
for individual conv layer. Thus, output pixels of the same
image plane will be in general convolved with different filters.  Note
that we do not explicitly force the kernel matrices to be identical, which
would recover the original convolution operation.

In this study, we simulate the RAPA ConvNet in order to validate the
effectiveness of different ways to distribute the compute among the
tiles and show that it is possible to achieve superior performance to
conventional ConvNets with the same kernel matrix sizes.  We further
prove analytically in a simplified model that for a random assignment
of compute to tiles, our architecture is indeed implicitly
regularized, such that tiles tend to learn similar kernel matrices.
Finally, we find that the RAPA ConvNet is actually more robust to
white-box adversarial attacks, since random assignment acts as a
``confidence stabilization'' mechanism that tends to balance
overconfident predictions.

\section{Convolution with replicated kernel matrices}
\label{sec:tile_methods}
Following common practice (e.g. \cite{chetlur2014cudnn}), the
convolution of a filter of size $\kh\times\kw$ over an input image of
size $\nh \times \nw \times \nc$ can be formulated as a matrix-matrix
multiplication between an $\np \times \k$ {\it im2col} matrix $\I$,
constructed by stacking all $\np$ (typically overlapping) image
patches $\a_i$ of size $\kh\times\kw\times\nc$ in rows of length
$\k=\kh\kw\nc$.  We can then write
$\I = \left(\a_1,\ldots,\a_\np\right)^T \equiv
\left(\a^T_i\right)_{i\in\{1,\ldots,\np\}}$.  The matrix $\I$ is then
multiplied by the $\k \times \ncout$ kernel matrix $\K$, where
$\ncout$ is the number of output channels (i.e. the number of
filters). The result $\M=\I\K$ is of size $\np\times\ncout$, and is
finally reshaped to a tensor with size
$\tilde{\nh}\times\tilde{\nw}\times \ncout $, to reflect the original
image content.

In most ConvNets, conv layers are alternated with some form
of pooling layers, that reduce the spatial size typically by a factor
of 2 (the pool stride)~\cite{gu2017recent}.  Thus, for the next
convolutional layer, $\np$ is reduced by a factor of 4 (square of the
pool stride).  On the other hand, because output channels become the
input channels to the following layer, the size of $\K$ changes as
well (see~\figref{compute_volume}).

\begin{figure}[t!]
  \centering
  \includegraphics[width=1\textwidth]{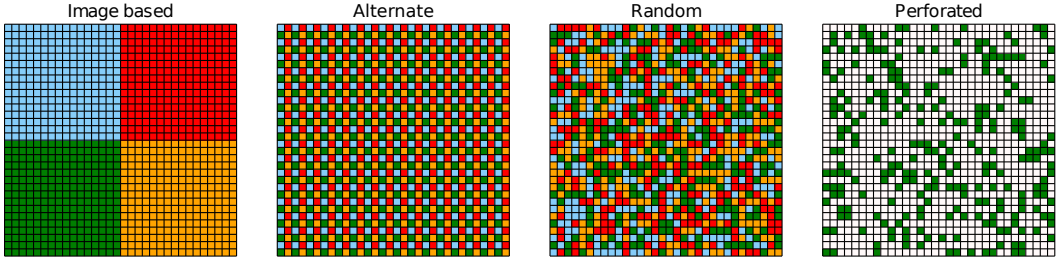}
  \caption{Illustrated is the output of a conv layer for different
    tiling schemes ($\ntiles=4$, $\ncout=1$).  Each output pixel might
    be computed with a kernel matrix from a different array tile (colors;
    white means zeros).  }
  \label{fig:tiling_illu}
\end{figure}

Our approach to parallelize the compute on analog arrays consists in
using $\ntiles$ kernel matrices $\K_j$ instead of just one $\K$ for a
given conv layer, and distributing the patches $\a_i$ equally among
them, so that at any given time $\ntiles$ matrix-vector products can
be processed in parallel.  Each of the $\np$ patches is assigned to
exactly one subset $S_j\subset \{1,\ldots,\np\}$ (all of roughly equal
size, $|S_j|\approx\np/\ntiles$), and the individual array tiles
effectively compute the sub-matrices
$\M_j = \I_j\K_j = \left(\a^T_{l}\right)_{l\in S_j}\K_j$.  How the
image patches are divided into the subsets $S_j$ is what we call
``tiling scheme'' (see below).

The final result is then obtained by re-ordering the rows according to
their original index. In summary, with $s_l=j$ if $l\in S_j$, we can
write
$\M_\text{tiled} =
\left(\a^T_{l}\K_{s_l}\right)_{l\in\{1,\ldots,\np\}}$.  Note that if
all $\K_j$ are identical, the tiled convolution trivially recovers the
original convolution.  If we assume that each kernel matrix $\K_j$
resides on a separate analog array tile, and all resulting $\I_j\K_j$
operations can be computed in parallel, the overall computation is
sped up by a factor of $\ntiles$ (neglecting the effort of the assignment,
since that can be done efficiently on the digital side of the mixed
analog-digital system).

However, if all $\K_j$ are learned independently and without explicit
synchronization (a prerequisite for embarrassingly parallel execution)
filters corresponding to the same output channel might in general be
non-identical, which implies that $\M_\text{tiled}\neq \M$.  Thus,
learning all $\K_j$ in parallel might negatively impact accuracy. In
the following, we test how different tiling schemes affect the overall
accuracy. We use the following schemes (compare to
\figref{tiling_illu}).

\paragraph{Image-based tiling}

This tiling scheme consists in collecting all patches that contain
pixels from a particular image region into a common subset $S_j$.  If
the image is a square with sides of length $n$ and the number of tiles
$\ntiles$ is a square number, $\ntiles=q^2$, the patch $\a_i$ centered
at pixel position $(x_i,y_i)$ with $x_i,y_i\in\{0,\ldots,n-1\}$ is
assigned to the subset $S_{s_i}$, with
$s_i = \left\lfloor \frac{q x_i}{n}\right\rfloor + q\left\lfloor
  \frac{q y_i}{n}\right\rfloor + 1 $.  Note that image patches at the
border will generally contain pixels from the neighboring regions.  We
thus call this scheme ``image w/overlap''.  Alternatively, the pixels
from other regions can be set to zero (as if padded in case of
separate sub-images), and we call this scheme ``image w/pad''.

\paragraph{Alternate tiling}
If the image is again a square and $\ntiles=q^2$, one could put image
patches that are neighboring to each other into different subsets, so
that neighboring image patches are assigned to alternate tiles. Specifically,
$ s_i = (x_i\mod q) + q\;(y_i\mod q) + 1$. This tiling is similar to
the ``tiled convolution'' approach suggested by~\cite{ngiam2010tiled}
as a way to improve the learning of larger rotational and
translational invariances within one convolutional layer.

\paragraph{Random tiling}
An alternative way of distributing $\np$ image patches onto $\ntiles$
kernel matrices, is to let the $S_j$ be a random partition of the set
$\{1,\ldots,\np\}$, with each of the $S_j$ having (roughly) the same
size.  We investigate two cases: one where the partition is drawn once
at the beginning and fixed the remainder (``random fixed''), and the
case where we sample a new partition for each train or test image
(``random'').

\begin{table}[t]
  \centering
  \caption{Best test (train) error [\%] for tiling schemes}
  \begin{tabular}{lccc}
    \toprule
    Tiling $\backslash$ Data & CIFAR-10 & SVHN & CIFAR-100  \\\hline
    \midrule
    no tiling & 18.85 \train{2.37}& 8.92 \train{1.96}& 47.99 \train{9.11}\\
    $\quad$ perforated & 30.79 \train{25.93}& 13.02 \train{15.52}& 63.44 \train{50.17}\\
    $\quad$ enlarged  & {\bf  17.75} \train{0.25}& 8.79 \train{0.71}& {\bf 46.91} \train{1.72}\\
    \midrule
    random [fixed] & 24.42 \train{3.86}& 11.28 \train{2.25}& 55.50 \train{23.72}\\
    random & {\bf 17.67} \train{5.81}& {\bf 7.10} \train{4.13}& 48.10 \train{15.57}\\
    image w/overlap & 24.52 \train{0.99}& 10.26 \train{3.01}& 53.22 \train{18.53}\\
    image w/pad & 25.86 \train{6.53}& 11.26 \train{6.06}& 54.24 \train{28.80}\\
    alternate & 21.02 \train{3.98}& 9.22 \train{2.99}& 52.08 \train{18.83}\\

    \bottomrule
  \end{tabular}

  \label{tab:tiling}
\end{table}

\paragraph{Perforated convolution}
An alternative way to speed up convolutions, is to simply train a
single kernel matrix with only a fraction $\np/\ntiles$ of the
data~\cite{figurnov2016perforatedcnns}.  As a result many output
pixels will have zero value.  Thus, in this scheme we randomly draw a
subset $S$ of $\np/\ntiles$ indices and set the rows for which
$i\notin S$ to $\boldsymbol{0}$, as described for
\cite{ngiam2010tiled}.  We resample $S$ for each image during training
and use all available image patches during testing.  Note that in this
scheme only a single kernel matrix is used.

\section{Network parameters used in the experiments}
\label{sec:params}

We perform a battery of proof of concept experiments using a small
standard ConvNet on 3 datasets: CIFAR-10,
CIFAR-100~\cite{krizhevsky2009learning}, and
SVHN~\cite{netzer2011reading}.  The network~\footnote{We used the
  ``Full'' network (except from changing the sigmoid activations to
  ReLu) from the Caffe examples in
  \url{https://github.com/BVLC/caffe/tree/master/examples/cifar10/}{}
} consists of 3 conv layers with kernel size $5\times 5$, and
intermediate pooling layers of stride 2.  We tried several options for
the first 2 pooling layers (see below), whereas the last pooling layer
is fixed to an average pooling.  Each conv layer is followed by
lateral response normalization, and the last conv layer is followed by
a fully connected layer.  We also use a very small weight decay
(0.0001 times the learning rate) and mini-batch of 10, train for
$>400$ epochs and report the minimal test and train errors.  The
learning rate $\lr$ is annealed in a step-wise manner every 25 epochs
with a factor $\lrgamma$, and is manually optimized for max-pooling on
CIFAR-10, then kept fixed for other datasets and pooling methods.  If
multiple runs on the datasets were made with different learning rate
settings, we report the best test error.  We found that $\lr=0.005$
and $\lrgamma=0.5$ for no tiling, and $\lr=0.05$ and $\lrgamma=0.75$
for tiling with $\ntiles=(16,4,1)$ tiles seemed to work best, although
different settings, e.g.\ $\lr=0.01$ and $\lrgamma=0.9$ yield mostly
similar results. Note that the number of updates is effectively
reduced per array tile, which can be in part compensated by increasing
the learning rate.  We additionally use a constant ``warm up'' period
of 1 or 5 epochs with a learning rate reduced by a factor of 50.

The output channel setting of the network is $32,32,64$ for the conv
layers, respectively.  Thus, for CIFAR-10 the network has 79328
weights (including biases) only in the conv layers.  For tiling with
$\ntiles=(16,4,1)$ tiles, the number of convolutional weights are
increased to 192704.  To compare this against a network of roughly the
same number of weights, we increase the number of channels for the
non-tiled network to $54,64,64$, which yields 193032 weights
(``enlarged'' network). However, note that for this larger network the
amount of compute is actually increased, whereas the amount of compute
of the tiled network is identical to the original smaller network.

For training we used standard stochastic gradient descent.  We use
moderate image augmentations (mirroring and brightness changes).  All
experiments are implemented in Facebook's Caffe2 framework (using
custom C++/CUDA operators, where necessary).

Finally, in addition to the usual pooling methods (max-pooling,
average-pooling and stochastic pooling, reviewed e.g.\ in
\cite{gu2017recent}), we also applied mixed pooling to get the
benefits of both max and average pooling. In particular, similar to
\cite{yu2014mixed}, we use a learnable combination of average and
max-pooling, with mixture parameters per channel $\alpha_k\in[0,1]$.
To enforce these parameter limits, we set
$\alpha_k\equiv\frac{1}{1+\exp^{\mu\beta_k}}$ and train the $\beta_k$
with $\mu=10$ fixed.  Initial values are $\beta_k = 2/\mu$ to ensured
a bias towards max-pooling, which works best on the datasets used
here.

  \begin{table}
    \caption{Best test (train) error [\%] for different pooling methods (CIFAR-10)}
    \label{tab:pooling}
    \centering
    \begin{tabular}{lccccc}
      \toprule
      Network  & no tiling  & no tiling, enlarged  &random & random  & random reduced \\
      Channel  & $(32,32,64)$ & $(54,64,64)$ & $(32,32,64)$ &  $(32,32,64)$ &  $(32,32,64)$\\
     $\ntiles$  & $(1,1,1)$    & $(1,1,1)$ &$(16,4,1)$ & $(16,4,1)$ & $(1,1,1)$  \\
     Performance    & single test & single test & single test & voting (5) & single test \\ 
      \midrule
      max pooling     &  18.93 \train{0.35}&  17.57 \train{0.04} &   17.67 \train{7.06} & 16.89 &19.31\\
      average       & 24.46 \train{4.29} &  23.28 \train{0.64}&  24.32 \train{7.64}& 24.23 &24.51\\
      mixed   &  18.19 \train{0.42}&  17.53 \train{0.04} &  17.37
                                                           \train{6.65}&
                                                                         {\bf 16.78} & 18.93 \\
      stochastic  & 20.09 \train{15.7}&  18.39 \train{11.02}&  21.15 \train{17.32}& 18.84 &21.19\\
      \bottomrule
    \end{tabular}
  \end{table}

\section{Results}
\label{sec:results}

\paragraph{Main experimental results}

Our aim here is to systematically quantify the relative impact of our
convolutional tiling architecture on performance, not to reach
state-of-the-art accuracy on the tested datasets.  We therefore
examine a relatively small standard ConvNet with 3 conv layers (see
\secref{params}).

As described, only the number $\np$ of input patches per layer
determines the run time on analog arrays. We thus divide the compute
of each conv layer onto $\ntiles$ array tiles, so that the number of
image patches per tile, $\np/\ntiles$, is constant. Since we have
$\np=(1024,256,64)$, we use $\ntiles=(16,4,1)$ tiles for the 3 conv
layers, respectively. Note that this architecture achieves perfect
load-balancing, because each tile in the network learns a separate
kernel matrix using $64$ image patches per image.

We tested the performance of this setup on the mentioned datasets with
and without tiling, and comparing different tiling schemes (see
\tabref{tiling}). The main results from these experiments are:
(1) ``Random'' tiling achieves the best performance among all tiling
schemes; (2) Across datasets, \emph{random tiling} actually beats the
regular ConvNet with no tiling; (3) Simply subsampling the input
images is not sufficient to explain the high performance of
\emph{random tiling}, since the \emph{perforated scheme} performed
poorly.

\paragraph{Filter similarity across tiles}
\renewcommand{\sim}{S}

Since replicated kernel matrices are trained independently, it is
interesting to examine the similarity of the filters at the end of
training. Note that only for identical filters across tiles, the
original convolution is recovered.

In general, two main factors tend to implicitly force kernel matrices
to become similar during training: (a) input similarity and (b)
error-signal similarity across tiles.  Indeed, for the random tiling
scheme, where the input distribution across tiles is identical on
average, different replicated filters might tend to be more similar,
but not for other tiling schemes.  Indeed, if we quantify the average
similarity $\sim$ of the learned filters across array tiles (computing
the average correlation coefficients between all pairs across tiles,
averaged over output channels) we find low values for all tiling
schemes trained with max-pooling ($S<0.01$), except for the random
tiling scheme.

\begin{figure}[t!]
  \includegraphics[width=1.15\textwidth,clip,trim=1.5cm 5.5cm 0cm 0.1cm]{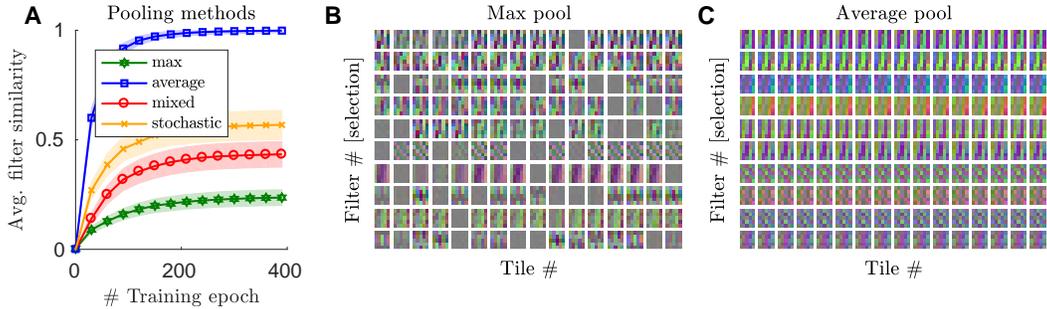}
  \caption{Similarity of learned kernel matrices $\K_j$ for the first
    convolution. (A) Similarity of $\K_j$ for random tiling and
    different pooling. (B) Selection of 10 out of 32 filters (rows of
    $\K_j$; reshaped) for all array tiles for max pooling . (C) Like B
    but for average pooling. }
  \label{fig:weight_sim_pooling}
\end{figure}

To investigate the effect of the error-signal, we further trained
random tiling networks with different pooling methods on CIFAR-10 (see
\tabref{pooling} for performance).  For instance, in the case of
average pooling, all tiles contributing to pixels in a pooling region
will receive the same error signal, whereas for max-pooling only one
output pixel per pooling region is selected and used to update the
corresponding tile.  We find that all pooling methods induce some
degree of similarity in case of random tiling ($S>0.1$; see
\figref{weight_sim_pooling}~B for example filters for max pooling).
We see the highest similarity for average pooling, where all tiles
learn almost identical filters ($S\approx 1$, see
\figref{weight_sim_pooling}~A and~C). However, average pooling gives
poor performance, suggesting that some diversity among replicated
kernel matrices might be advantageous.  A good trade-off between
similarity and performance can thus be obtained by using a learnable
mixture between max and average pooling (\figref{weight_sim_pooling}A
and \tabref{pooling} mixed pooling).

\paragraph{Comparison with larger model and predictions based on majority vote}
Our experiments show that random tiling matches or even outperforms
the original network (see \tabref{tiling} and \tabref{pooling}).
However, since replicating kernel matrices onto multiple tiles
effectively increases the number of free parameters in the network (by
about a factor of 2.5, see \secref{params}), it seems fair to compare
the performance of the tiled network with a network with a similar
number of free parameters arranged in conventional fashion.  When we
do that by increasing the number of channels of a non tiled network
(which however increases the amount of compute; see \secref{params}),
we do indeed find that this enlarged network achieves a performance
comparable to the random tiling network (see \tabref{tiling} and
\tabref{pooling}).

It is worth noticing that the performance of the random tiling network
in \tabref{tiling} is obtained by sampling only one random assignment
of patches to tiles during test.  For each test image, we can instead
generate multiple predictions, each generated by a different random
assignment, and take as final output the majority vote of all
predictions (similarly e.g.\ to \cite{graham2014fractional}).  We test
this majority vote over 5 predictions, and see a performance gain of
roughly 1\% accuracy for the random tiling network, which then
outperforms even the enlarged network with adjusted number of
parameters (see~\tabref{pooling} second last column). Note, however,
that there is no performance gain in case of average pooling, where
filters become almost identical (\figref{weight_sim_pooling}A),
indicating an additional benefit of diversity among filter replica at
test time.

\paragraph{Reduction of tiled network to the original architecture}

It might be problematic for certain applications to retain multiple
kernel matrices per conv layer.  Thus, one might want to
recover the original network, after benefiting from the training
speedup of the tiled network. If the filters are very similar (as with
average pooling) just taking a kernel matrix of any tile recovers the
original convolution and the performance of the original network
(see~\tabref{pooling} last column).

One way to reduce the tiled model for mixed or max-pooling, is to
select among all replica the filters that most often ``wins'' the
maximum pooling on the training set.  These can then be combined to
form a single kernel matrix.  An alternative simpler way is to just
select across tiles the filter with the highest norm, since that indicates
a filter that is more often used and updated, and therefore less
subject to the weight decay penalty.

We tested this last reduction technique and found that the reduced
network's performance is only slightly worse than the original network
with conventional training ($<0.75$\% for max/mixed pooling,
see~\tabref{pooling}), indicating no need for retraining. However,
note, that reducing the network to the original architecture also
removes the benefits of accelerated run time on analog arrays,
the performance gain by majority voting, and the robustness to adversarial
attacks (investigated below).

\paragraph{Theoretical analysis: Implicit regularization of random tiling}
\label{theory}
\newcommand{\vct}[1]{\boldsymbol{#1}} 
\newcommand{\s}{{\vct{s}}}
\newcommand{\thetas}{\vct{\theta}^{\s}}
\newcommand{\thetam}{\vct{\bar{\theta}}}
\newcommand{\R}{R(\{\thetas\})}

It is rather intriguing that our random tiling scheme achieves a
performance that is comparable or even better than the standard
ConvNet.  One might have expected that as many as 16 replicated kernel
matrices for one conv layer would have incurred overfitting.  However,
empirically we see that random tiling actually tends to display less
overfitting than the standard ConvNet.  For example in
\tabref{pooling} (first row) we see that the standard ConvNet (no
tiling) achieves a test error of 18.93\% with a training error close to
zero, while random tiling has a better test error rate of 17.67\% with
higher training error (7.06\%). In this section, we give a formal
explanation of this phenomenon and show in a simplified model, a
fully-connected logistic regression model, that replicating an
architecture's parameters over multiple ``tiles'' that are randomly
sampled during training acts as an implicit regularization that helps
to avoid overfitting.

A logistic regression is a conditional distribution over outputs
$y\in \{0,1\}$ given an input vector $\vct{x}\in \mathbb{R}^d$ and a
set of paramters $\vct{\theta} \in \mathbb{R}^d$.  The exponential
family distribution form of the logistic regression is
$p(y\vert \vct{x}, \vct{\theta}) =
\exp\left(y~\vct{x}\cdot\vct{\theta} -
  A(\vct{x}\cdot\vct{\theta})\right)$,
where $A(z)\equiv-\log(1-\sigma(z))$ and
$\sigma(z)\equiv(1+\exp(-z))^{-1}$ is the logistic function. Note that
this expression is equivalent to the more common form
$p(y=1\vert \vct{x}, \vct{\theta}) =
\sigma(\vct{x}\cdot\vct{\theta})$. Training a logistic regression
consists in finding parameters that minimize the empirical negative
log-likelihood,
$l_{\vct{x},y}(\vct{\theta}) = -\log p(y\vert \vct{x}, \vct{\theta})$,
over a given set of $N$ training examples $(\vct{x}^i, y^i)$,
resulting in the minimization of the loss:
$L(\vct{\theta}) = \sum_{i=1}^N l_{\vct{x}^i,y^i}(\vct{\theta})$.

We model random tiling by assuming that every parameter $\theta_l$ is
being replicated over $\ntiles$ tiles.  Correspondingly, every time
$\theta_l$ is being accessed, a parameter $\theta_l^{s_l}$ with $s_l$
randomly sampled in $\{1,\ldots,\ntiles\}$ is retrieved. We write
$\thetas \equiv (\theta_l^{s_l})_l$ and $\s\equiv (s_l)_l$.  As a
result training can be expressed as the minimization of the average
loss,
$\left\langle L(\thetas)\right\rangle_\s = \sum_{i=1}^N \left\langle
  l_{\vct{x}^i,y^i}(\thetas)\right\rangle_\s$, where the angular
brackets $\langle\cdot\rangle_\s$ indicate averaging over the process
of randomly sampling every parameter $\theta_l$ from a tile $s_l$.
With the above, we get
$\left\langle
  L(\thetas)\right\rangle_\s=-\sum_{i=1}^N\left(y^i~\vct{x}^i\cdot\vct{\bar{\theta}}
  -\left\langle
    A\left(\vct{x}^i\cdot\thetas\right)\right\rangle_\s\right) =
L(\thetam)+\R$, where $\thetam$ is the vector whose components are the
parameters averaged across tiles, i.e.\
$\bar{\theta}_l = \langle\theta_l^{s_l}\rangle_\s$, and
\begin{equation*}
\R = \sum_{i=1}^N\left(\left\langle A\left(\vct{x}^i\cdot\thetas\right)\right\rangle_\s - A\left(\vct{x}^i\cdot\thetam\right)\right).
\end{equation*}
The term $\R$ that falls out of this calculation has the role
of a regularizer, since it does not depend on the labels $y^i$.  In a
sense, it acts as an additional cost penalizing the deviations of the
replicated parameters $\thetas$ from their average value
$\thetam$ across tiles.  This tendency of the replicated
parameters to move towards the mean counteracts the entropic pressure
that training through stochastic gradient descent puts on the replica
to move away from each other (see e.g.\ \cite{Zhang2018}), therefore
reducing the effective number of parameters.  This implicit
regularization effect explains why, despite the apparent
over-parametrization due to replicating the parameters over tiles, our
architecture does not seem to overfit more than its standard
counterpart.  It also explains the tendency of the tiles to
synchronize causing the filters to become similar
(\figref{weight_sim_pooling}).

\paragraph{Robustness against adversarial examples}

We can gain further intuition on the role of the regularizer
$\R$ by developing its first term as a Taylor series up
to second order around $\vct{x}^i\cdot\vct{\bar{\theta}}$, analogously
to what is done in \cite{Bishop1995, Wager2013}. 
This results in:
\begin{equation*}
\label{eq:confidence_stabilization}
\R
\approx \frac{1}{2} \sum_{i=1}^N A''\left(\vct{x}^i\cdot\vct{\bar{\theta}}\right) \sum_l (x_l^i)^2~\text{Var}_\s(\theta_l^{s_l})
= \frac{1}{2} \sum_{i=1}^N p_i(1-p_i) \sum_l (x_l^i)^2~\text{Var}_\s(\theta_l^{s_l}),
\end{equation*}
where $\text{Var}_\s(\theta_l^{s_l})$ is the variance of the parameter
$\theta_l$ across tiles, and
$p_i=\sigma\left(\vct{x}^i\cdot\vct{\bar{\theta}}\right)$ is the
predicted probability that $y^i=1$ when considering the parameter mean
$\thetam$.  This penalty $\R$ can be interpreted as trying to
compensate for high-confidence predictions (for which the term
$p_i(1-p_i)$ is small) by diminishing the pressure on
$\text{Var}_\s(\theta_l^{s_l})$ to be small.  As a result, samples
$\vct{x}^i$'s for which the prediction will tend to be confident will
be multiplied by weights $\theta_l$ that will display a relatively
large variability across replica, which in turn will tend to reduce
the degree of confidence.

This ``confidence stabilization'' effect raises the intriguing
possibility that random tiling mitigates the weaknesses due to a model
excessively high prediction confidence.  The efficacy of
\emph{adversarial examples}, i.e.\ samples obtained with small
perturbations resulting in intentional high-confidence
misclassifications, is such a type of weakness that plagues several
machine learning models \cite{goodfellow2014explaining}.  Our
analysis, suggests that random tiling should help immunize a model
against this type of attacks, by preventing the model from being
fooled with high confidence.

\begin{figure}[t]
   \centering
   \includegraphics[width=0.4\textwidth,clip,trim=0.9cm 8.6cm 9.4cm 1.0cm]{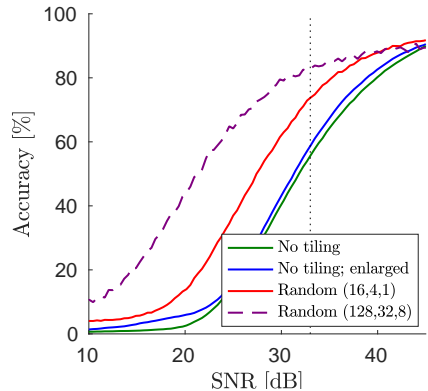}
   \caption{Tiling improves robustness to adversarial examples. }
   \label{fig:adversarial}
\end{figure}

We verify the theoretical prediction that random tiling increases the
robustness to adversarial samples by using the Fast Gradient Sign
Method (FSGM)~\cite{goodfellow2014explaining} to attack a network
trained on CIFAR-10 with max-pooling (see performance results in
\tabref{pooling}).  In particular, we computed the accuracy drop from
all correctly classified images in the test set, due to a perturbation
by noise in the direction of the signed error
gradient~\cite{goodfellow2014explaining} (with strength $\epsilon$).
Following \cite{cisse2017parseval}, we computed the drop in accuracy
as a function of the signal-to-noise ratio resulting from adversarial
noise (see \figref{adversarial}).
At a noise level corresponding to the threshold of human perception,
$\epsilon\approx 33$ (according to \cite{cisse2017parseval}), we find
that random tiling reduces the gap to perfect adversarial robustness
by around $41$\%.  In comparison, other learning methods, such as
\cite{cisse2017parseval} or enhancing training examples with
adversarial gradients~\cite{goodfellow2014explaining} reduces the gap
on CIFAR-10 by around $6\%$ and $54$\%, respectively (using their
baseline, compare to~\cite{cisse2017parseval}, Table~1).  While the
networks used here are not the same as those used in
\cite{cisse2017parseval}, our results still suggest that random tiling
significantly improves robustness, with no loss in performance or
extra training examples.

A strategy to further improve robustness is to increase the number of
tiles in the random tiling network.  If we set $\ntiles=(128,32,8)$
the network still trains fine, reaching a test error of $16.83$\% on
CIFAR-10, which is similar to the $\ntiles=(16,4,1)$ tiled network
(within 500 epochs; max-pool; majority vote of 9 tests; compare to
\tabref{pooling}).  However, now robustness to adversarial attacks is
significantly improved, reaching an accuracy of $83.97$\% for
$\epsilon\approx 33$ (see \figref{adversarial}; dashed line), which
translates to a reduction of the gap to perfect robustness by $64$\%.
Note that, although the $\ntiles=(128,32,8)$ network has now about
$20$ times more convolutional weights than the original non-tiled
network, it trains well and does not overfit (training error 15\%)
and, neglecting peripheral costs and assuming parallel execution of
all analog array tiles in a layer, would execute a training epoch
$128$ times faster than the original network.

\section{Discussion}

We here propose a way how to modify ConvNets, so that they map more
favorably onto upcoming mixed analog-digital hardware.  Interestingly,
we find that using multiple independently trained kernel matrices per
convolution instead of a single one, and randomly dividing the compute
among them, yields no loss in accuracy. Our architecture has the added
advantage that, executed on parallel analog arrays, it would in
principle run $\ntiles$-times faster than conv layers run on an
individual analog array.  We show that random assignment regularizes
the training, avoiding overfitting, and, additionally, increases the
robustness towards adversarial attacks.

We here studied and validated the principles of our architecture in a
small standard ConvNet. However, we expect the tiling architecture to
be applicable also to larger ConvNets
(e.g.~\cite{krizhevsky2012imagenet}), because they generally
successively reduce the spatial size with depth through
pooling~\cite{gu2017recent} and thus have a similar pattern of
the amount of compute per layer as our example network
(\figref{compute_volume}). For instance, an efficient tiling of the
architecture in \cite{krizhevsky2012imagenet} would be
$\ntiles=(17,4,1,1,1)$. This would achieve perfect load-balancing
across the 5 conv layers on analog arrays.  Note that if set-up in
this way, the whole network (including the fully connected layers) can
additionally be pipelined across image
batches~\cite{ben2018demystifying}, because the duration of
computation would be identical for each of the conv layers
(irrespective of the different filter sizes and numbers of channels).

There are many different approaches to accelerate deep learning using
current hardware~\cite{ben2018demystifying}. Our approach is motivated
by the constraints of mixed-analog digital hardware to emphasize its
advantages. In our tiling approach, although the total amount of
compute in the network is kept constant (contrary to e.g. methods that
perforate the loop~\cite{figurnov2016perforatedcnns}, or use low-rank
approximations or low precision weights, reviewed in
\cite{gu2017recent}), the number of updates per weight is nevertheless
reduced, which might generally affect learning curves. Importantly,
however, this does not seem to have an impact on the number of
training epochs needed to achieve a performance close to the best
performance of conventional networks. In fact, the random tiling
network (with majority vote) reaches a test error of 19\% (mixed
pooling, see \tabref{pooling}) after 85 epochs versus 82 for the
original network.  Admittedly, if one is instead interested in
reaching the superior performance of the random tiling network, one
would typically need to add additional training time. To what degree
the added training time could be reduced by heterogeneous learning
rates across the tiled network, is subject of future research.

Finally, another interesting research direction is how the performance
of RAPA ConvNets could be further improved by increasing the
convolution filter size or the number of filters per layer.
Remarkably, this type of modifications, which are generally avoided on
GPUs for reasons of efficiency, would not alter the overall run time
on upcoming mixed analog-digital hardware technology.

{\footnotesize
\bibliographystyle{unsrt}
\bibliography{nips}
}

\end{document}